\documentstyle[12pt]{dubna98}
\newcommand{\co}{\; \; ,}

\newcommand{\scs}{\co \;}
\newcommand{\per}{ \; .}

\newcommand{\calleff}{{\cal L}_{\mbox{\small{eff}}}}

\newcommand{\ed}{\end{document}}
\newcommand{\be}{\begin{equation}}
\newcommand{\ee}{\end{equation}}
\newcommand{\bea}{\begin{eqnarray}}
\newcommand{\eea}{\end{eqnarray}}

\begin{document}
\begin{flushright}
BUTP-98/22
\end{flushright}
\vskip2cm
\begin{center}
{\bf $\pi\pi$ SCATTERING AT LOW ENERGY: STATUS REPORT\footnote{
Talk given at the International
Workshop "Hadronic Atoms and Positronium in the Standard Model", Dubna, Russia,
 May 26-31, 1998. To appear  in the Proceedings.
}\\}
\vspace*{1cm}
J\"URG GASSER\\
{\it Institute of Theoretical Physics,\\}
{\it University of Bern, Sidlerstrasse 5, 3012 Bern, Switzerland}

\vskip.8cm

September 1998

\end{center}
\vspace*{.8cm}
\begin{abstract}
{\small I discuss the status of ongoing work  to
determine numerically the elastic $\pi\pi$ scattering amplitude at
order $p^6$ in  the framework of chiral perturbation theory.}
\end{abstract}

\vspace*{.2cm}
\section{Introduction}
The interplay between theoretical  and experimental
aspects of elastic $\pi\pi$ scattering
is illustrated in figure 1.
 On the theoretical side, Weinberg's calculation\cite{weinberg66}
of the scattering amplitude  at
leading order in the
low--energy expansion gives for the isospin zero $S$--wave scattering
 length the value
$a_{l=0}^{I=0}=0.16$ in units of the charged pion
mass. This  differs from the experimentally determined
value\cite{rosselet}
$a_0^0=0.26 \pm 0.05$ by two standard deviations. The one--loop
calculation\cite{glplb} enhances the leading order term to  $a_0^0=0.20\pm
0.01$ -- the correction goes in the
right direction, but the result is still on the low side as far as the
 present experimental value is concerned.
 To decide about agreement/disagreement between theory and experiment,
 one should i)  evaluate
the scattering lengths in the theoretical framework at order $p^6$, and ii)
determine them more precisely experimentally. My talk was concerned with the
the former issue, indicated by the double arrow in the left column of
the figure.

On the {\it experimental} side, several  attempts are underway to
 improve our knowledge of the threshold parameters.
 The most
promising ones among them are semileptonic $K_{l4}$ decays with improved
 statistics (E865\cite{e865} and KLOE\cite{kloe}),
 and the  measurement of the pionium lifetime
 (DIRAC\cite{dirac}) that will allow one to directly determine
 the combination $|a_0^0-a_0^2|$ of
 $S$-wave scattering lengths.
It was one of the aims of the workshop to discuss  the precise relation
between
the lifetime of the pionium atom and the  $\pi\pi$ scattering
lengths --
  I refer the
interested reader to the numerous  contributions to this workshop
for details.
 In addition, Po\v{c}ani\'c\cite{pocanic} has presented
a nice review of  the determination of scattering lengths from
 hadronic processes.

As for {\it theory}, I will describe in the following sections
our current effort\cite{acglw}
 to determine the scattering amplitude
 numerically at order $p^6$ in the framework of chiral perturbation theory
 (CHPT).
 As the title of my talk indicates,
 this work is still in progress -- therefore, I will give an overview
 of what is going, omitting any  details.
 For related work, see Refs.\cite{royrelated,knecht,wanders,atw}.
\begin{figure}[h]
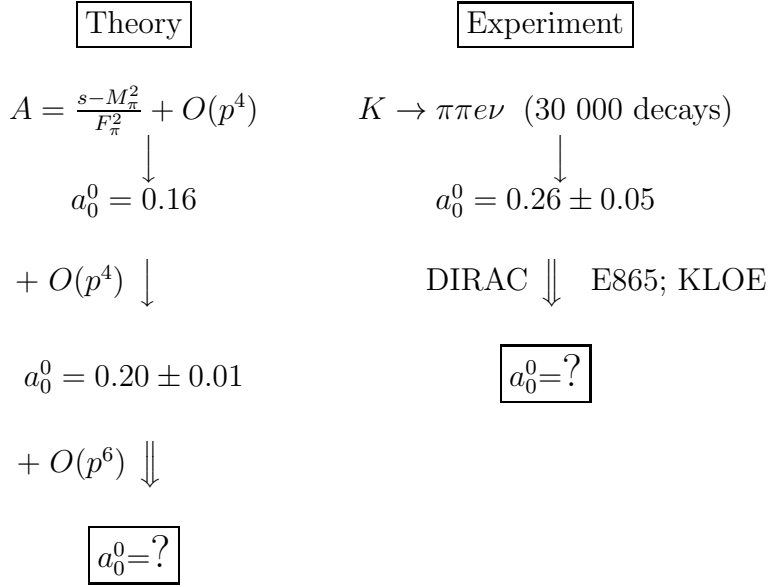

 \begin{center}
$
\hspace{1cm}\begin{array}{ccc}
\framebox{Theory}&\hspace{0cm}&\framebox{Experiment}
\\&&\\
A=\frac{s-M_\pi^2}{F_\pi^2}+O(p^4)&& K\rightarrow \pi\pi e \nu
\;\; (\mbox{30 000 decays})\\
\left.
\begin{array}{c} \\ \end{array}
\right\downarrow &&
\left.
\begin{array}{c} \\ \end{array}
\right\downarrow
\\
a_0^0=0.16 && a_0^0=0.26\pm 0.05
\\&&\\
\hspace{-1.5cm}\left.
\begin{array}{c}+\;O(p^4)  \\ \end{array}\right\downarrow&&
\hspace{1.3cm}\left.\begin{array}{c} {\mbox{DIRAC}} \\
\end{array}\right\Downarrow\begin{array}{c}
  {\mbox{ E865; KLOE}}\\ \end{array}
\\&&\\
a_0^0=0.20\pm 0.01&&\framebox{$a_0^0$=\mbox{\Large{?}}}
\\&&\\
\hspace{-1cm}\left.
\begin{array}{c}+\;O(p^6) \\ \end{array}\right\Downarrow
\begin{array}{c}  \\\end{array}&&
\\
\\
\framebox{$a_0^0$=\mbox{\Large{?}}}&&
\end{array}
$
\end{center}
 \caption{
Progress in the determination of the elastic $\pi\pi$ scattering
amplitude. References are provided in the text.}
\label{fig1}
\end{figure}
\section{Chiral perturbation theory}
To set  notation, I briefly describe the framework  used\cite{acglw} to
evaluate the $\pi\pi$ scattering amplitude. I
consider the isospin symmetry limit where the two lightest quark masses are
equal,
$m_u=m_d\neq 0$, and where the electromagnetic coupling is set to zero,
 $\alpha_{QED}=0$.
 I refer to M\"uller's\cite{mueller} contribution for a discussion of  isospin
violating effects in hadronic  amplitudes.
 The effective lagrangian that describes
the interaction of  pions is given by a string of terms,
$
\calleff={\cal{L}}_2 +\hbar {\cal{L}}_4 +\hbar^2{\cal{L}}_6+\cdots\scs
$
where ${\cal{L}}_{n}$ contains $m_1$ derivatives of the pion fields and $m_2$
quark mass matrices, with $m_1+2m_2=n$  (here, I consider the standard
counting rules\cite{weinberg79,glannpb}--
 see section 7  for  a generalization thereof\cite{gchpt}). The low--energy
expansion corresponds to an expansion of the scattering matrix elements in
powers of $\hbar$.

In the following, the low--energy constants (LEC's) hidden in the effective
lagrangians ${\cal{L}}_n$ play an important role. In ${\cal{L}}_2$, there are
two
 of them, the pion decay constant in the chiral limit (F) and the
parameter $B$, which are related to the condensate by
$
F^2B=-\langle 0|\bar{u}u|0\rangle$. In the loop expansion, these two
parameters can be expressed in terms of the
physical pion decay constant $F_\pi \simeq 92.4$ MeV and of the pion
mass, $M_\pi = 139.57$ MeV.
The $\pi\pi$ scattering amplitude contains, in the two--loop approximation,
 in addition several LEC's
ocurring in ${\cal{L}}_{4}$ and in ${\cal{L}}_6$,
\bea
\begin{array}{ll}
\left.\begin{array}{l}
 {\cal{L}}_2:F_\pi, M_\pi\\
 {\cal{L}}_4:\bar{l}_1,\bar{l}_2,\bar{l}_3,\bar{l}_4\\
 {\cal{L}}_6: \bar{r}_1,\ldots, \bar{r}_6\end{array}\right\}&{\mbox{occur in}}
\;\pi\pi\rightarrow\pi\pi\; {\mbox{ (two--loop approximation)}}.\end{array}
\label{eq2.1}
 \eea
These LEC's are not determined by chiral symmetry -- they are, however, in
principle calculable in QCD\cite{rebbi}.

\section{$\pi\pi\rightarrow\pi\pi$ in CHPT}
Elastic $\pi\pi$ scattering
is described by a single Lorentz invariant amplitude $A(s,t,u)$, that
depends on the standard
Mandelstam variables $s,t,u$ (I use the same notation as
Sainio\cite{sainio}-- I refer the reader to his contribution for details).
Here, we are concerned
with the  loop expansion of the scattering amplitude,
\bea
A(s,t,u)=\left\{\begin{array}{ccccccc}\\
A_2&+&A_4&+&A_6& +&O(p^8)\co \\
\uparrow&&\uparrow&&\uparrow&&\\
{\mbox{tree}}&&{\mbox{1 loop}}&&{\mbox{2 loops}}&&\\
\end{array}\right.\label{eq3.1}
\eea
where  $A_n$ is of order $p^n$. In the following, I denote with
the symbol $A^\chi$ the first three terms in (\ref{eq3.1}),
\be
A^\chi=A_2+A_4+A_6\per
\ee
A dispersive evaluation of $A^\chi$ has been performed in Ref.\cite{knecht}.
That  calculation is not sufficient for
the present purpose, because several tadpole diagrams that occur in the
loop expansion have not been evaluated in that work.
 What is needed for  the analysis outlined below is the
 {\it complete two--loop expression} of $A^\chi$.
 The relevant calculation  has been performed in
Ref.\cite{bcegs1}.
The explicit form of the amplitude is  not needed here, and I refer the
interested reader to Sainio's contribution\cite{sainio}
for explicit formulae.

        For the following discussion, it is  useful to bring the amplitude into
a
canonical form.
First, one makes use of the fact that $A^\chi$ can be expressed\cite{sternw3}
 in terms of
three
functions $W_{0,1,2}$, that depend  on a single kinematical variable
($s$,$t$ or $u$) and have only a right
hand cut.
Their imaginary part is given by the imaginary parts of the three lowest
partial
waves $t_l^I$,e.g.,
${\mbox{Im}}W_0=\mbox{Im}t^{0}_{0}$, etc.
 We now write a four times subtracted dispersion relation
for $W_0$,
\be
W_0=a_0+a_1s+a_2s^2+a_3s^3+\frac{s^4}{\pi}\int_{4M_\pi^2}^\infty
\frac{dx}{x^4}\frac{\mbox{Im}t_0^0(x)}{x-s}\per
\ee
and similarly for $W_{1,2}$. Inserting these representations into the
expression for
 $A^\chi$ gives
\bea
A^\chi&=&P^\chi+A_{int}^\chi\per\nonumber\\
P^\chi&=&\beta_1^\chi+\beta_2^\chi s+\beta_3^\chi s^2+\beta_4^\chi (t-u)^2
+\beta_5^\chi s^3+
\beta_6^\chi s(t-u)^2\per\label{eq3.9}
\eea
The coefficients $\beta_i^\chi$ contain the
low--energy constants (\ref{eq2.1}),
\bea
\beta_n^\chi=\beta_n^\chi(\bar{l}_1,\ldots
 ,\bar{l}_4;\bar{r}_1,\ldots,\bar{r}_6).
\eea
Here and in the following, I drop all dependence on $F_\pi$ and $M_\pi$, whose
numerical value is taken from experiment. It is the aim of the
enterprise to work out the consequences of the
requirement that $A^\chi$ is a good approximation of the
true amplitude at low
energy. This requirement allows one in particular to work out the
two $S$--wave
scattering lengths $a^0_0$ and $a^2_0$, provided that we can
determine the LEC's ocurring in the chiral amplitude.

\section{Effective couplings}
The chiral  amplitude $A^\chi$ contains two sets of LEC's:
\begin{itemize}
\item[\framebox{set 1}] In this set we put the four LEC's
\bea
\framebox{$\bar{l}_1,\bar{l}_2,\bar{r}_5,\bar{r}_6$}
\label{eq4.1}
\eea
that are related to the momentum and scattering angle dependence of
the amplitude:
 the polynomial $P^\chi$ contains
terms of the form
$
\bar{l}_1s(t-u)^2, \bar{l}_2(t-u)^2, \bar{r}_5s^3,
\bar{r}_6 s(t-u)^2.
$
 As a result of this, these couplings can be determined from data above
$\simeq 800$ MeV, see below.

\vskip .2cm

\item[\framebox{set 2}]
This set  assembles the LEC's whose effect disappears in the chiral
limit, because they are multiplied with the square of the pion mass.
Examples are the terms
$
M_\pi^6\bar{r}_1, M_\pi^4s \bar{r}_2
$
 in $P^\chi$. As nature does not allow us to vary the pion mass in the
laboratory,
 we need input from outside $\pi\pi$ scattering to determine these couplings.
 I denote these with
the symbol $\Theta$,
\be
\Theta: \framebox{$\bar{l}_{3,4};\bar{r}_{1,2,3,4}$} \per
\ee
Using large--$N_c$ arguments, the constant $\bar{l}_3$ can be
obtained from the chiral expansion of the ratio
$M_K^2/M_\pi^2$ \cite{glannpb},
 whereas $\bar{l}_4$ is related to the scalar radius of the
pion\cite{glannpb,bct}.
Finally, the effect of $\bar{r}_{1,2,3,4}$
is  suppressed by  powers of the pion mass -- a
rough estimate like the one provided in Ref.\cite{bcegs1}
(based on resonance saturation) thus suffices.
\end{itemize}

\section{Matching}
{}From the above discussion, we find that the coefficients $\beta_n^\chi$ in
the polynomial $P^\chi$ of the chiral amplitude  (\ref{eq3.9}) are of the
functional form
\be
\beta_n^\chi=\beta_n^\chi(\bar{l}_1,\bar{l}_2,\bar{r}_5,\bar{r}_6;
\Theta)\per
\ee
In the following, I consider  the couplings $\Theta$  as given.

Turning now to experiment, I write the corresponding amplitude
$A^{exp}$ as
\bea
A^{exp}=P^{exp}+A_{int}^{exp \label{eq5.2}}
\eea
where $P^{exp}$ has the same functional form as $P^\chi$, with
$\beta_n^\chi\rightarrow \beta_n^{exp}$. As we will see in the following
section,
 the Roy equations\cite{roy} fix these coefficients in terms of data above
 $\simeq$ 800 MeV and of the scattering lengths $a_0^0,a_0^2$,
\be
P^{exp}\leftrightarrow \beta_n^{exp}(a_0^0,a_0^2)\per
\ee
 Furthermore, the amplitude
$A_{int}^{exp}$ coincides with $A_{int}^\chi$ up to terms of order $p^8$.
We now require that the chiral amplitude agrees with the experimental one
near threshold , as a result of which one obtains the relation
\bea
\framebox{$
\begin{array}{c}
P^\chi (s,t,u)=P^{exp}(s,t,u)\\
\mbox{matching condition}\end{array}$}\per\label{eq5.3}
\eea
 In terms of the coefficients $\beta_n$, this matching condition amounts
to the six relations
\be
\beta_n^\chi
(\bar{l}_1,\bar{l}_2,\bar{r}_5,\bar{r}_6;\Theta)=\beta_n^{exp}(a_0^0,a_0^2)\; ;
\; n=1,\ldots , 6
\ee
for six unknowns. Solving for these, we
can determine
the quantities
\be
\framebox{$a_0^0,a_0^2; \bar{l}_1,\bar{l}_2,\bar{r}_5,\bar{r}_6$}\per
\label{eq5.6}
\ee
The remaining threshold parameters may be obtained from the Wanders sum
rules\cite{wanderssum,atw}.

\section{Roy equations}
It remains to show that indeed the polynomial coefficients $\beta_n^{exp}$
are determined in terms of the two $S$--wave scattering
lengths $a_0^0$, $a_0^2$ and of data above 800 MeV.

For this purpose, one consider the partial waves $t_l^I$.
 As shown by Roy\cite{roy}, their real and imaginary parts are
 related through
\bea
\mbox{Re}t_l^I=c_l^I(s)+\sum_{I'=0}^2\sum_{l'=0}^\infty P
\hspace{-4mm}\int_{4M_\pi^2}^\infty
dx
K^{II'}_ {ll'}(s,x){\mbox{Im}}t_{l'}^{I'}(x)
\label{eq6.1}
\eea
in the region $-28M_\pi^2<s<60 M_\pi^2$. Here, the quantities $K^{II'}_{ll'}$
are known kernels, whereas the subtraction polynomial $c_l^I$ contains the two
scattering lengths $a_0^0, a_0^2$ as the only free parameters. Note that the
relation
 (\ref{eq6.1}) is linear in the partial wave amplitudes -- it is therefore also
true in ordinary perturbation theory, order by order in the coupling
 constant (modulo subtractions). Next, we observe that, at low energy, the most
important contribution to the imaginary parts stems from the $S$-- and
$P$--waves, because the higher waves are suppressed,
$
\mbox{Im}{t_l^I}=O(p^8)\; ; \; l =2,3,\ldots \per
$
For this reason, it is  useful to perform the splitting\cite{wmr}
\bea
A^{exp}=A_{SP}+A_R\co\label{eq6.2}
\eea
where the crossing symmetric amplitude  $A_{SP}$   contains the exact
$S$-- and $P$--wave absorptive parts, whereas $A_R$ has only absorptive
parts from $l=2,3,\ldots$ waves. Therefore, at low energies, the latter
 contribution  can be approximated by  a polynomial in
 $s,t$ and $u$.
This allows one finally to write $A^{exp}$ in the
form (\ref{eq5.2}), where $P^{exp}$ is determined from $A_R$  and through the
imaginary parts of the three lowest partial waves.  I now show how one may
determine these imaginary parts from experimental information above 800 MeV.

\vskip.5cm

Using (\ref{eq6.2}) and expanding $A_R$ as  described, the partial wave
relations (\ref{eq6.1}) become for the $S$-- and $P$-- waves
\bea
\mbox{Re}{t}_m=c_m+\sum_{n=0}^{2} P \hspace{-4mm}\int_{4M_\pi^2}^\infty dx
\overline{K}_{mn}(s,x)\mbox{Im}t_n(x) +d_m(s)\; ; \; m=0,1,2
\label{eq6.1.1}
\eea
where
\bea
(t_0^0,t_1^1,t_0^2)\rightarrow (t_0,t_1,t_2)\; ; \;
(c_0^0,c_1^1,c_0^2)\rightarrow (c_0,c_1,c_2)\per
\eea
The
kernels are $\overline{K}_{00}$ = $K_{00}^{00}$, etc. Finally,  the so--called
driving
terms $d_m$ are obtained from the amplitude $A_R$.
The crucial step\cite{roy} is to invoke unitarity at this stage,
\be
t_m=\frac{1}{2i}\frac{1}{(1-4M_\pi^2/s)^{1/2}}
\left(\eta_m\exp{2i\delta_m}-1\right)\per
\ee
 We now assume that
\begin{itemize}
\item
the elasticities $\eta_m$ are known for all $s$
\item
 the $\delta_m$ are given above a cutoff energy $E_{cut}\simeq $ 800 MeV
\item
the driving terms are known below $E_{cut}$.
\end{itemize}
At given $a_0^0,a_0^2$, the equations (\ref{eq6.1.1}) then become a system of
 nonlinear integral equations
 for the unknown $\delta_m$ in the interval $2M_\pi < \sqrt{s} < E_{cut}$.
This system is known as 'Roy equations' in the literature. Solving
these\footnote{For $E_{cut} \simeq 800$ MeV, the solution is
unique\cite{wandersunique}.},
one obtains the $S,P$ partial waves at low energies and thus the
polynomial $P^{exp}$, as promised.

\vskip.5cm

\section{Generalized chiral perturbation theory}
The above framework assumes the standard scenario of  chiral counting, where
the condensate is the leading order parameter, considered a quantity
of order one.
In recent years, an alternative picture\cite{zerocondensate} has been
implemented\cite{gchpt} in an effective
lagrangian framework. In that picture, the condensate may be small or even
vanishing. Many predictions are lost in this case, a prominent
one being the Gell--Mann--Okubo mass relation, that does not
have a natural explanation in this framework. Furthermore, the
scattering length $a_0^0$ cannot be predicted either  -- it
is a free parameter, related to the
size of the condensate. The philosophy of this approach is to let experiment
 tell the size of $a_0^0, a_0^2$, which allows one then to determine
 the size of the condensate. I refer the interested reader to Stern's talk in
Mainz\cite{sternmainz} for further details and for references.

It is not easy to distinguish phenomenologically a small condensate from
the standard case with the present precision of low--energy
experiments\cite{sternmainz}.
It is, however, expected that new  precise
measurements of the $\pi\pi$ amplitude at low energy will shed more light on
the issue.

On the theoretical side,  a recent interesting investigation by Knecht and
de Rafael\cite{derafael}
has shown that, in vector--like gauge theories like QCD, and at large
values
of $N_c$, the ordering pattern of narrow vector and axial--vector states
 is correlated with the size of possible local order parameters
 of chiral symmetry breaking.
The authors find that, from a duality point of view, the option of a
vanishing condensate seems unnatural. I refer the reader to their work
for more details.
Recent lattice calculations\cite{aoki,vladikas} do not support
a small condensate either. Indeed, Ecker\cite{ecker} has compared the results
of Ref.\cite{aoki} on the quark mass dependence of the meson masses with the
 predictions of standard and generalized CHPT and
concludes that ``\ldots lattice QCD is incompatible with a
small quark condensate''.
The authors of Ref.\cite{vladikas} have made a considerable effort
to pin down systematic uncertainties in a direct evaluation
 of
 $\langle 0|\bar{q}q|0\rangle$ on the lattice in the quenched approximation.
They come up with  the standard value,
 with remarkably small error bars.

\section{Summary}

\vskip.3cm

\framebox{Status}

\vskip.3cm
\begin{itemize}
\item
The analytic form of the two--loop amplitude in standard
CHPT is known\cite{bcegs1}.
\item We can numerically construct the solution of the system
(\ref{eq6.1.1}) for given scattering lengths, input data and driving
terms\cite{acglw}.
\item The determination of the
 uncertainties in the
final values of the parameters (\ref{eq5.6}), in particular in the scattering
lengths, is in progress\cite{acglw}.
\item
For related work, see
Refs.\cite{royrelated,knecht,wanders,atw}.

\end{itemize}

\vskip.3cm

\framebox{Outlook}

\vskip.3cm
\begin{itemize}
\item
GCHPT\cite{gchpt} provides a framework where the condensate may
be small or even
 vanishing.
\item
In this scenario, the isospin zero $S$--wave scattering
length $a_0^0$
can be large, in contrast to the standard case presented here, and in contrast
to
lattice calculations\cite{aoki,vladikas,ecker} in the quenched approximation.
\item
We all hope  that DIRAC\cite{dirac}, E865\cite{e865}, KLOE\cite{kloe} and
CHAOS\cite{chaos}
will soon provide additional information  on the issue from the
experimental side.
\end{itemize}

\vskip.5cm

\section*{Acknowledgements}
I was very much impressed by the most pleasant working and recreational
atmosphere that the
organizers had provided for us, and I wish to thank them cordially for
 their hospitality.
This work has been supported in part by Schweizerischer Nationalfonds,
by TMR, EC-Contract No. ERBFMRX-CT980169 (EURODA$\Phi$NE), and by
Bundesamt f\"ur Bildung und Wissenschaft,
BBW-Contract No. 97.0131.


\begin{thebibliography}\\
\bibitem{weinberg66}
S. Weinberg, {\it Phys. Rev. Lett.} {\bf 17} (1966) 616.

\bibitem{rosselet}
L. Rosselet et al., {\it Phys. Rev.} {\bf D15} (1977) 574.

\bibitem{glplb}
J. Gasser and H. Leutwyler, {\it Phys. Lett.} {\bf B125} (1983) 325.

\bibitem{e865}
Experiment E865 at Brookhaven AGS; S. Pislak, talk given at the
 $K_{l4}$ meeting in Bern, Switzerland, June 29--30, 1998.

\bibitem{kloe}
M. Baillargeon and P.J. Franzini, in Ref.\cite{dafneii}, p. 413;
 J. Lee--Franzini, in Ref.\cite{dafneii}, p. 761;
P.J. Franzini, in Ref.\cite{workmit}, p. 117.

\bibitem{dafneii}
L. Maiani, G. Pancheri and N. Paver (eds.), The Second DA$\Phi$NE
Physics Handbook (INFN, Frascati, 1995).

\bibitem{workmit}
A.M. Bernstein and B.R. Holstein (eds.), Chiral Dynamics: Theory and
Experiment, Proceedings of the Workshop held at MIT, Cambridge, MA, USA,
25--29 July 1994 (Springer, Berlin and Heidelberg, 1995).

\bibitem{dirac}
B. Adeva et al., Proposal to the SPSLC: Lifetime measurement of $\pi^+\pi^-$
atoms to test low--energy QCD predictions,  CERN/SPSLC/P 284, December 15,
1994; L. Nemenov, J. Schacher, these proceedings.

\bibitem{pocanic}
D.  Po\v{c}ani\'c, these proceedings.

\bibitem{acglw}
B. Ananthanarayan et al., work in progress.



\bibitem{royrelated}
M.R. Pennington and S.D. Protopopescu, {\it Phys. Rev.} {\bf D7} (1973) 1429;
2591;
J.L. Basdevant, C.D. Froggatt and J.L. Petersen, {\it Nucl. Phys.} {\bf B72}
 (1974) 413;
C.D. Froggatt and  J.L. Petersen, {\it Nucl. Phys.} {\bf B91} (1975) 454;
 ibid. {\bf B104} (1976) 186 (E); ibid. {\bf B129} (1977) 89;
O.O. Patarakin, V.N. Tikhonov and K.N. Mukhin, {\it Nucl. Phys.} {\bf A598}
(1996) 335;
B. Ananthanarayan and P. B\"uttiker,
{\it Phys. Rev.} {\bf D54} (1996) 1125; 5501.


\bibitem{knecht}
M. Knecht, B. Moussallam, J. Stern and N.H. Fuchs, {\it Nucl. Phys. }{\bf B457}
(1995) 513; ibid.
 {\bf B471} (1996) 445.

\bibitem{wanders}
G. Wanders, {\it Phys. Rev.}  {\bf D56} (1997) 4328;
{\it Helv. Phys. Acta} {\bf 70} (1997) 287.

\bibitem{atw}
B. Ananthanarayan, D. Toublan and G. Wanders, {\it Phys. Rev.}
{\bf D51} (1995) 1093;
ibid. {\bf D53}
 (1996) 2362;
 B. Ananthanarayan and P. B\"uttiker, {\it Phys. Lett.} {\bf B415}
 (1997) 402.

\bibitem{mueller}
G. M\"uller, these proceedings.

\bibitem{weinberg79}
S. Weinberg, {\it Physica} {\bf 96A} (1979) 327.

\bibitem{glannpb}
J. Gasser and H. Leutwyler, {\it Ann. Phys. (N.Y.)}  {\bf 158} (1984) 142;
{\it Nucl. Phys.} {\bf B250} (1985) 465.

\bibitem{gchpt}
N.H. Fuchs, H. Sazdjian and J. Stern, {\it Phys. Lett.} {\bf B269} (1991) 183;
J. Stern, H. Sazdjian and N.H. Fuchs, {\it Phys. Rev.} {\bf D47} (1993) 3814;
M. Knecht and J. Stern, in Ref.\cite{dafneii}, p. 169, and references
cited therein.

\bibitem{rebbi}
S. Myint and C. Rebbi, {\it Nucl. Phys.} {\bf B421} (1994) 2416;
A.R. Levi, V. Lubicz and C. Rebbi,
 {\it Phys. Rev.} {\bf D56} (1997) 1101;
{\it Nucl. Phys. Proc. Suppl.} {\bf 53} (1997) 275.

\bibitem{sainio}
M.E. Sainio, these proceedings.

\bibitem{bcegs1}
J. Bijnens et al.,
{\it Phys. Lett.} {\bf B374} (1996) 210; {\it Nucl. Phys.} {\bf B508} (1997)
 263;
ibid. {\bf B517} (1998) 639 (E).

\bibitem{sternw3}
J. Stern, H. Sazdjian and N.H. Fuchs, in Ref.\cite{gchpt}.


\bibitem{bct}
J. Bijnens, G. Colangelo and P. Talavera, {\it J. High Energy Phys}. {\bf 05}
(1998) 014 (hep-ph/9805389).

\bibitem{roy}
S.M. Roy, {\it Phys. Lett.} {\bf 36B} (1971) 353; {\it Helv. Phys. Acta} {\bf
63}
 (1990) 627.

\bibitem{wanderssum}
G. Wanders, {\it Helv. Phys. Acta} {\bf 39} (1966) 228.

\bibitem{wmr}
G. Mahoux, S.M. Roy and G. Wanders, {\it Nucl. Phys.} {\bf B70} (1974) 297.

\bibitem{wandersunique}
G. Wanders, to appear.

\bibitem{zerocondensate}
M.D. Scadron and H.F. Jones, {\it Phys. Rev.} {\bf D10} (1974) 967;
H. Sazdjian and J. Stern, {\it Nucl. Phys. } {\bf B94} (1975) 163;
N.H. Fuchs, {\it Phys. Rev.} {\bf D14} (1976) 1709.

\bibitem {sternmainz}
J. Stern,
{\it Light quark masses and condensates in QCD}, hep-ph/9712438, to appear in:
 Proceedings of the Chiral Dynamics Workshop in Mainz, Germany,
 September 1-5, 1997.

\bibitem{aoki}
S. Aoki et al. (CP-PACS), {\it Nucl. Phys. Proc. Suppl.} {\bf 60A} (1998) 14.

\bibitem{vladikas}
L. Giusti et al., {\it The QCD Chiral Condensate from the Lattice},
Edinburgh preprint 98/10 (hep-lat/9807014).


\bibitem{ecker}
G. Ecker, {\it Chiral symmetry}, hep-ph/9805500,
 to appear in:
       Proc. of 37. Internationale Universit\"atswochen f\"ur Kern- und
Teilchenphysik,
       Schladming, Austria, Feb. 28 - March 7, 1998.

\bibitem{derafael}
 M. Knecht and E. de Rafael,
     {\it Phys. Lett.} {\bf B424} (1998) 335.


\bibitem{chaos}
M.E.  Sevior, {\it Determination of the $\pi^\pm p\rightarrow \pi^\pm\pi^+n$
 Cross--Section Near Threshold},
 in: Working group on $\pi\pi$ and $\pi N$ interactions, p.11
(hep-ph/9711361),
 to appear in:
 Proceedings of the Chiral Dynamics Workshop in Mainz, Germany,
 September 1-5, 1997.

\end{thebibliography}
\end{document}